\documentclass[conference]{IEEEtran}
\IEEEoverridecommandlockouts
\usepackage{cite}
\usepackage{amsmath,amssymb,amsfonts}
\usepackage{graphicx}
\usepackage{textcomp}
\usepackage{makecell}
\usepackage{xcolor}
\usepackage{threeparttable}
\usepackage{multirow}
\usepackage{subfigure}

\makeatletter
\newif\if@restonecol
\makeatother

\usepackage[linesnumbered,ruled]{algorithm2e}
\usepackage{algpseudocode}

\def\BibTeX{{\rm B\kern-.05em{\sc i\kern-.025em b}\kern-.08em
    T\kern-.1667em\lower.7ex\hbox{E}\kern-.125emX}}
\begin{document}

\title{Hardware Accelerator for Multi-Head Attention and Position-Wise Feed-Forward in the Transformer\\
}
\author{Siyuan~Lu,
        Meiqi~Wang,	
		Shuang~Liang,
        Jun~Lin,
        and Zhongfeng~Wang\\
		\IEEEauthorblockA{School of Electronic Science and Engineering, Nanjing University, Nanjing, China\\	
			Email: \{sylu,~mqwang,~sliang\}@smail.nju.edu.cn, \{jlin,~zfwang\}@nju.edu.cn}
\vspace{-3em}
\thanks{This work was supported by the National Natural Science Foundation of China under Grant 61604068, the Fundamental Research Funds for the Central Universities under Grant 021014380065, the Key Research Plan of Jiangsu Province of China under Grant BE2019003-4. (Corresponding authors: Jun Lin; Zhongfeng Wang.)
}
}

\maketitle

\begin{abstract}
Designing hardware accelerators for deep neural networks (DNNs) has been much desired. Nonetheless, most of these existing accelerators are built for either convolutional neural networks (CNNs) or recurrent neural networks (RNNs).
Recently, the Transformer model is replacing the RNN in the natural language processing (NLP) area.
However, because of intensive matrix computations and complicated data flow being involved, the hardware design for the Transformer model has never been reported.
In this paper, we propose the first hardware accelerator for two key components, i.e.,~the multi-head attention (MHA) ResBlock and the position-wise feed-forward network (FFN) ResBlock, which are the two most complex layers in the Transformer.
Firstly, an efficient method is introduced to partition the huge matrices in the Transformer, allowing the two ResBlocks to
share most of the hardware resources.
Secondly, the computation flow is well designed to ensure the high hardware utilization of the systolic array, which is the biggest module in our design.
Thirdly, complicated nonlinear functions are highly optimized to further reduce the hardware complexity and also the latency of the entire system.
Our design is coded using hardware description language (HDL) and evaluated on a Xilinx FPGA. Compared with the implementation on GPU with the same setting, the proposed design demonstrates a speed-up of 14.6$\times$ in the MHA ResBlock, and 3.4$\times$ in the FFN ResBlock, respectively. Therefore, this work lays a good foundation for building efficient hardware accelerators for multiple Transformer networks.
\end{abstract}


\section{Introduction}

Recurrent neural networks (RNNs), long-short memory (LSTM)\cite{hochreiter1997long}, and gated recurrent (GRU)\cite{chung2014empirical}, used to be the best solutions in the natural language processing (NLP) area. This situation was changed when the Transformer model\cite{vaswani2017attention} was invented in 2017, which outperforms previous RNN models in multiple tasks.
By avoiding the recurrent calculations and taking full advantage of the attention mechanism, the Transformer and Transformer-based pre-trained language models (such as BERT\cite{devlin2019bert}, ALBERT\cite{lan2019albert}, T5\cite{raffel2019exploring}, ERINE\cite{sun2019ernie}, and structBERT\cite{wang2019structbert}) have achieved state-of-the-art accuracy in various NLP tasks.

In spite of making great progress in relative fields, the high computation complexity and huge memory requirements of these powerful  Transformer networks are making them hard to be operated in mobile devices or embedded systems.
More and more researchers are paying attention to this problem, and one way to solve it is through model compression\cite{ganesh2020compressing}. Several techniques have been used to compress these networks, including data quantization\cite{bhandare2019efficient}, pruning, knowledge distillation and Architecture-Invariant Compression (AIC)\cite{lan2019albert}.

Recently, building FPGA or ASIC hardware accelerators for deep neural networks (DNNs) has achieved great success in both academic and industrial societies, which makes us believe that designing efficient hardware architectures for these Transformer networks must be an important topic as well.
By implementing them on hardware platforms, the inference systems of many NLP applications, such as machine translation, question answering, and sentiment analysis, are able to achieve higher speed or lower power consumption or both. However, intense matrix computations, complicated data flow, and complex non-linear functions are making it hard to design efficient hardware architecture for the Transformer.
To the best of our knowledge, we are the first to propose a specific hardware accelerator for the Transformer.
In the open literature, the $A^3$\cite{ham20203} is the only hardware architecture for accelerating the attention mechanism in various neural networks, which is not specifically designed for the Transformer.

As mentioned in \cite{vaswani2017attention} and \cite{lan2019albert}, most of the trainable parameters and the computations are in the multi-head attention (MHA) ResBlock and the position-wise feed-forward network (FFN) ResBlock, which is discussed by Section \uppercase\expandafter{\romannumeral2} in detail.
In this work, we design a reconfigurable hardware architecture based on systolic array (SA) for the MHA ResBlock and the FFN ResBlock, which are the two most complex layers in the Transformer.

Main contributions of this work can be summarized as follows:
\begin{itemize}
\item We provide an efficient method to partition the huge matrices in the Transformer, which allows the MHA ResBlock and the FFN ResBlock to share most of the hardware resources.
\item We propose the first hardware architecture design which can complete the calculations for both these two ResBlocks. To ensure the high hardware utilization of the SA, which is the biggest module in our design, the computation flow is well designed.
\item Two most complicated nonlinear functions, including the scaled masked-softmax and the layer normalization, are highly optimized to become more hardware-friendly. As the ``bottle-neck'' in the proposed architecture, the latency of layer normalization is reduced as much as possible.
\end{itemize}

After quantizing the Transformer base model in \cite{vaswani2017attention} (distinguished from the Transformer big model) with 8-bit integers (INT8), we also evaluate our design on the Xilinx xcvu13p-fhga2104-3-e FPGA, when the max sequence length (denoted as $s$) is equal to 64 and the batch size is equal to 1.
The hardware experimental results demonstrate a speed-up of 14.6$\times$ in the MHA ResBlock, and a speed-up of 3.4$\times$ in the FFN ResBlock, compared to a GPU implementation on an NVIDIA V100.

The rest of this paper is organized as follows. Section \uppercase\expandafter{\romannumeral2} gives a brief review of the Transformer networks, and explains the importance of accelerating the MHA ResBlock and the FFN ResBlock.
Section \uppercase\expandafter{\romannumeral3} presents the method of matrix partitioning.
Section \uppercase\expandafter{\romannumeral4} describes the proposed hardware architecture. Experimental results are given in Section \uppercase\expandafter{\romannumeral5}. Section \uppercase\expandafter{\romannumeral6}
concludes this paper.

\section{Background and Motivation}

\subsection{The Model Architecture of the Transformer}

\begin{figure}[]
    \centering
    \includegraphics[scale=0.35]{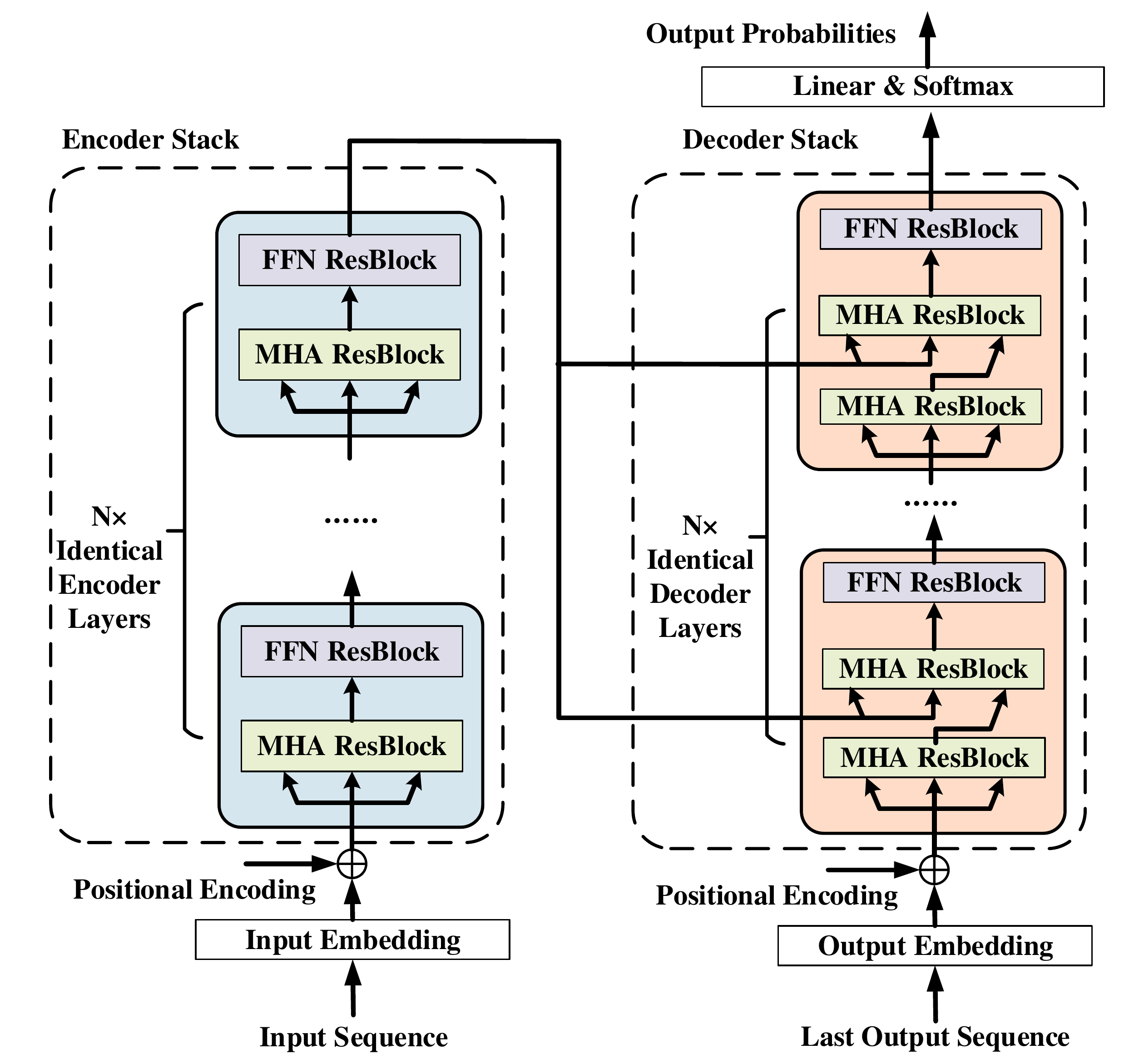}
    \vspace{-0.8cm}
    \caption{The model architecture of the Transformer.}
    \label{fig:Transmodel}
    \vspace{-0.6cm}
\end{figure}

The model architecture of the Transformer is described in Fig.~\ref{fig:Transmodel}, containing an encoder stack and a decoder stack. Notice that most of the trainable parameters and the computations are in these two stacks, and other components beside the stacks such as the embedding layers and the softmax output layer have not been taken into account by this work. As is shown in Fig.~\ref{fig:Transmodel}, all the encoder layers and the decoder layers are composed of two kinds of ResBlocks, the MHA ResBlock and the FFN ResBlock.

\begin{figure}[]
    \centering
    \includegraphics[scale=0.34]{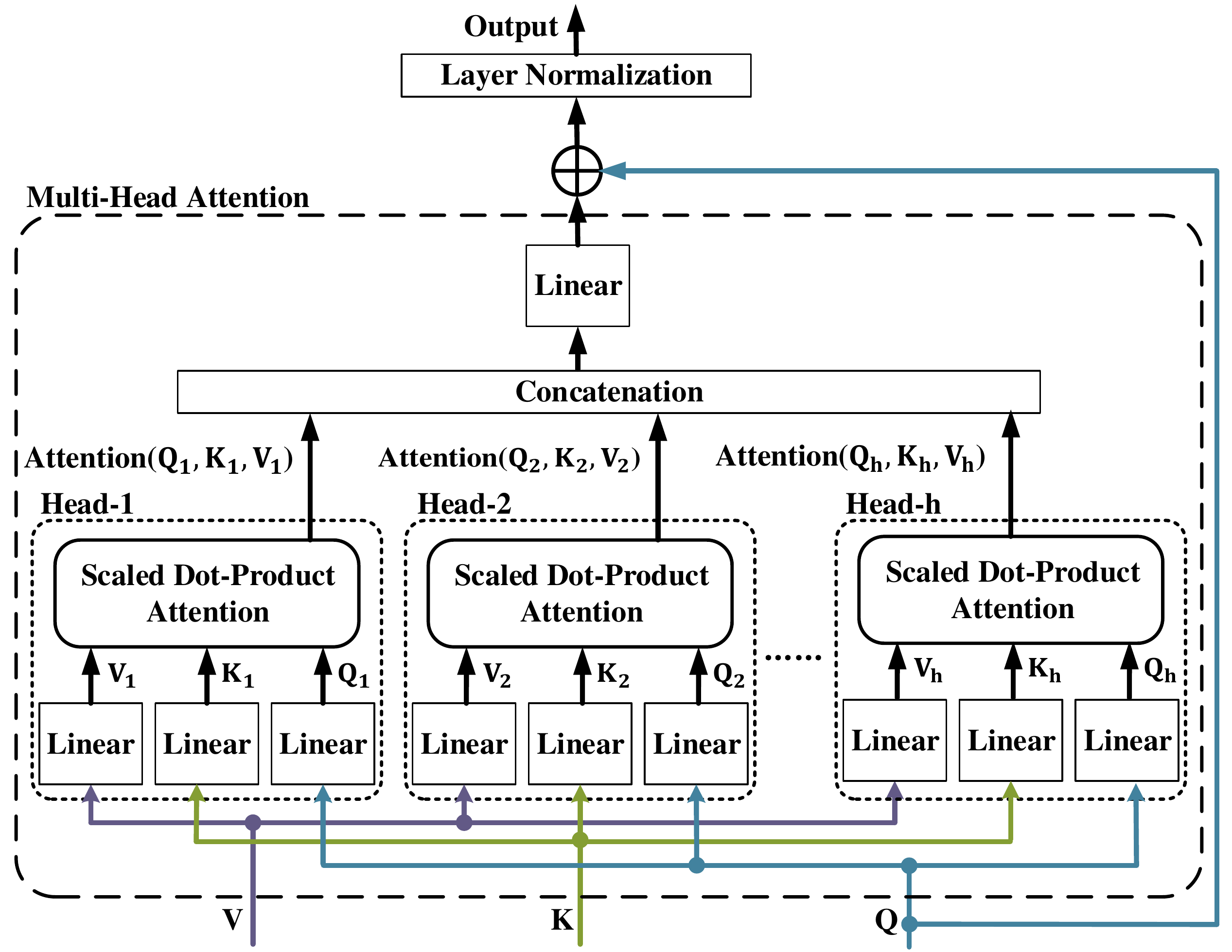}
    \vspace{-0.8cm}
    \caption{The structure of the MHA ResBlock.}
    \label{fig:MHA ResBlock}
\vspace{-0.6cm}
\end{figure}
Fig.~\ref{fig:MHA ResBlock} shows the structure of the MHA ResBlock. An MHA ResBlock has $h$ ``Attention Heads'', and the input of each Head is the same as the input of the ResBlock, including three tensors: V (values), K (keys), and Q (queries).
The Scaled Dot-Product Attention function in the MHA can be expressed as follows:
\begin{equation}
Attention(Q_i,K_i,V_i)=softmax(Mask(\frac{Q_iK_i^T}{\sqrt{d_k}}))V_i.
\end{equation}

The Mask operation is used to mask out all values in the input of the softmax corresponding to illegal connections, and the parameter $d_k$, which is equal to 64 in both the Transformer base model and the Transformer big model. The parameter $h$ is equal to 8 in the base model, or equal to 16 in the big model.

The FFN ResBlock contains a fully connected feed-forward network, consisting of two linear sublayers and a ReLU activation between them:
\begin{equation}
\begin{split}
\begin{aligned}
FFN(x)=ReLU(xW_1+b_1)W_2+b_2,\\
FFN\_ResBlock(x)=LayerNorm(x+FFN(x)).
\end{aligned}
\end{split}
\end{equation}

\subsection{Transformer-Based Pre-Trained Models}
An important pre-trained model is Bidirectional Encoder Representations from Transformers (BERT).
Analyses in \cite{ganesh2020compressing} also point out that, the MHA and the FFN ResBlocks still occupy most of the storage space and have the highest numbers of FLOPs.


\begin{figure*}[!htp]
    \subfigure[The MHA ResBlock.]{
        \label{MHA_Matrix}
        \includegraphics[scale=0.35]{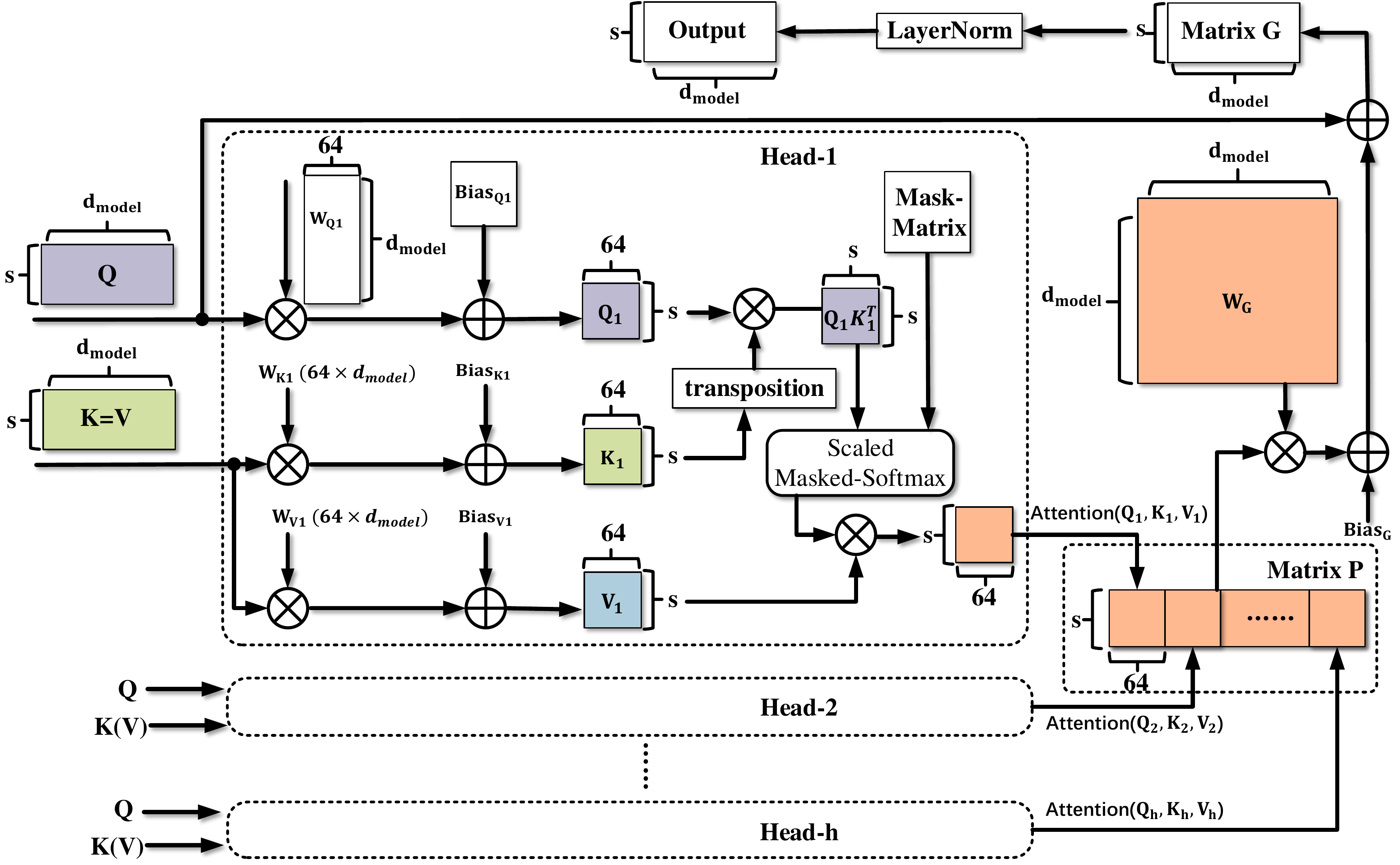}}
    \hspace{0.1in}
    \subfigure[The FFN ResBlock.]{
        \label{FFN_Matrix}
        \includegraphics[scale=0.35]{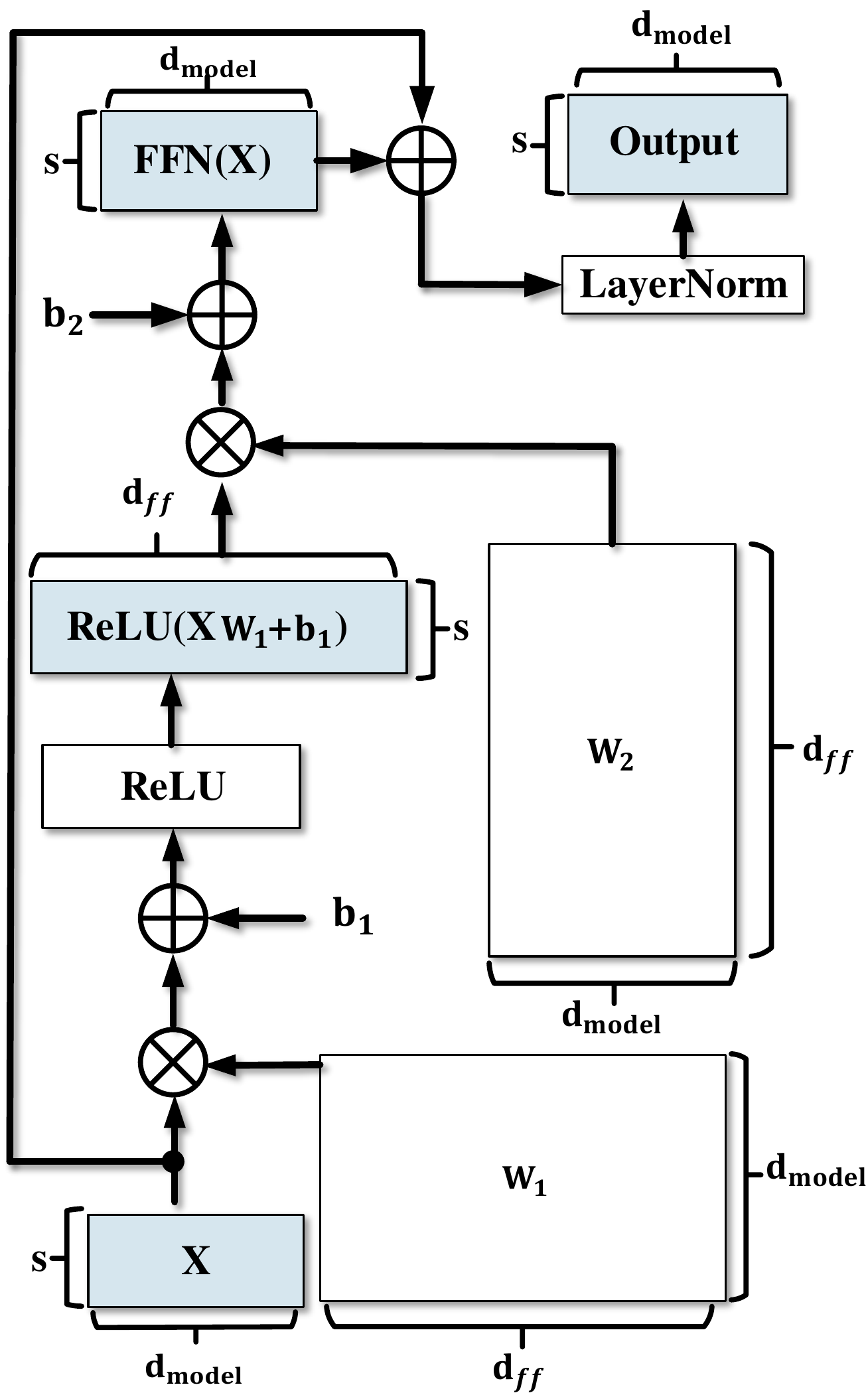}}
    \vspace{-0.3cm}
    \caption{Matrix Operations in the MHA and the FFN ResBlocks. Note that all the multiply operations marked in this figure are dealing with cross products.
     }
    \label{fig:Matrix}
    \vspace{-0.5cm}
\end{figure*}

The General Language Understanding Evaluation (GLUE) benchmark \cite{wang2018glue} is a collection of diverse natural language understanding tasks. Recently, many Transformer-based pre-trained models have obtained top placements on the GLUE score list. Most of these models, such as T5\cite{raffel2019exploring}, ERINE\cite{sun2019ernie}, and structBERT\cite{wang2019structbert}, have very similar structure to the BERT. These facts all prove the necessity of designing efficient hardware accelerators for the MHA and the FFN ResBlocks, which are two commonly used structures in these models.

\section{Partitioning Matrices in the FFN and the MHA}
Considering the characteristics of the Transformer architecture, we believe that the proposed hardware accelerator should be able to accelerate not only the MHA ResBlock, but also the FFN ResBlock.
To make sure that the MHA ResBlock and the FFN ResBlock can reuse the hardware resources, we first analyze these two ResBlocks from the perspective of matrix operations, and then give a method to partition the matrices so that all the general matrix-matrix multiplications (GEMMs) can be done with one and the same systolic array (SA), the size of which is limited to $s\times64$.

Assuming that the input of the FFN is called X, the shape of the tensor X is the same as Q (one of the input tensors of the MHA), which is [$batch\_size, seq\_len\_q, d_{model}$]. Additionally, Fig.~\ref{fig:Transmodel} shows that the tensor K is always equal to the tensor V, the shape of which is [$batch\_size, seq\_len\_v, d_{model}$]. In normal circumstances, $seq\_len\_q$ is equal to $seq\_len\_v$, so the shape of all these four tensors can be expressed as [$batch\_size, s, d_{model}$].
Supposing that the batch size is equal to 1, the computations of these two ResBlocks can be considered sets of matrix operations, which are represented in Fig.~\ref{fig:Matrix}.
Obviously, an $s\times64$ SA can support all the matrix multiplications in the Linear sublayers of all the Heads. However, how to complete other multiplications between larger matrices, including $P\times W_G$, $X\times W_1$, and $ReLU(XW_1+b_1)\times W_2$, is another important issue to be considered.

\begin{table}[htbp]

    \caption{Variations on the Transformer and the BERT architectures.} 
    \centering
    \begin{threeparttable}
    \begin{tabular}{p{2.8cm}|p{1.5cm}|p{1.5cm}|p{1cm}}
    \hline
    \hline
     & \makecell[tl]{$d_{model}$} & \makecell[tl]{$d_{ff}$} & \makecell[tl]{$h$}\\ 
    \hline 
    \hline
    Transformer-base  & 512  & 2048 & 8\\
    \hline
    Transformer-big & 1024 & 4096 & 16\\
    \hline
    $BERT_{BASE}$ & 768 & 3072 & 12\\
    \hline
    $BERT_{LARGE}$ & 1024 & 4096 & 16\\
    \hline
    \hline
    \end{tabular}
    \end{threeparttable}
    \label{Trans_variables}
    \vspace{-0.1cm}
    \end{table}

\begin{figure}[]
\vspace{-0.5cm}
    \centering
    \includegraphics[scale=0.35]{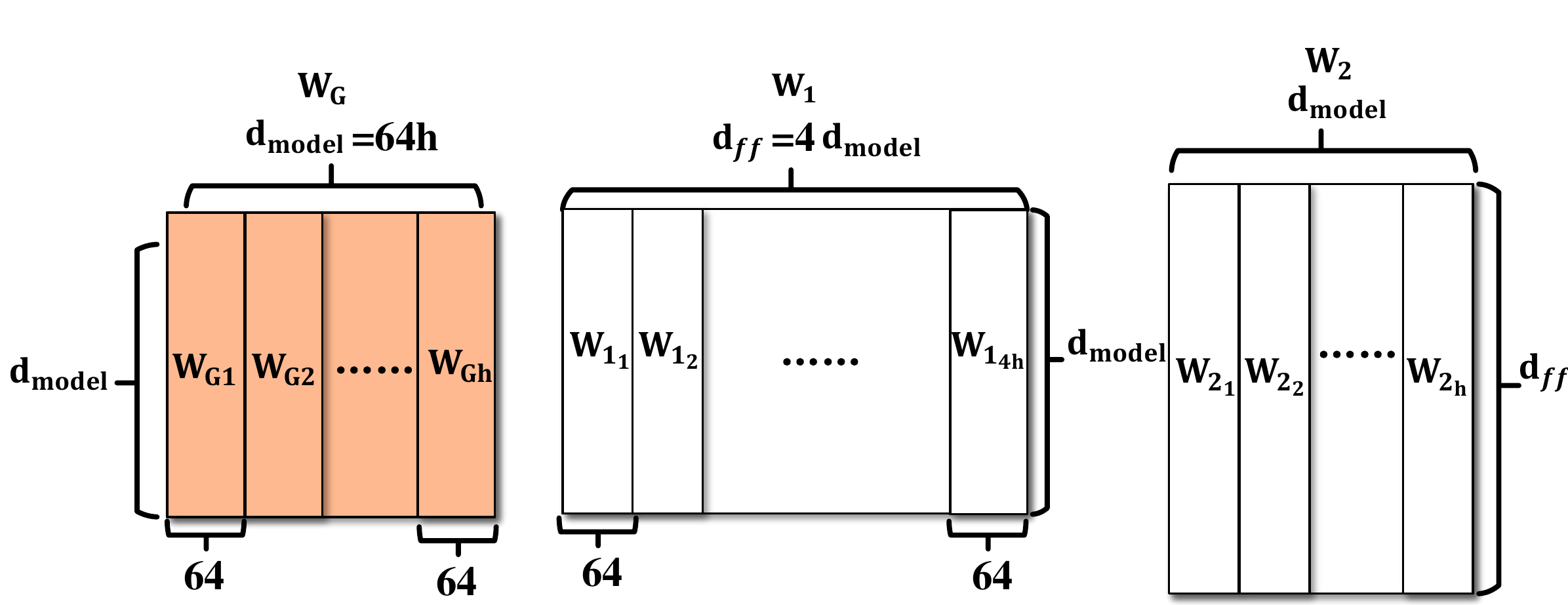}
    \caption{Partition $W_G$, $W_1$, and $W_2$.}
    \label{fig:Partition}
\vspace{-0.6cm}
\end{figure}

Table \ref{Trans_variables} shows that in these Transformer networks, we all have $d_{model}=64h$, and $d_{ff}=4d_{model}=256h$. On the basis of this pattern, the three large weight matrices $W_G$, $W_1$, and $W_2$ can be partitioned as shown in Fig.~\ref{fig:Partition}. Thus, most of the GEMMs can be done with an $s\times64$ SA.

The only one left is the operation of $Q_i\times K_i^T$ in each Head of the MHA. 
The ratio of the number of multiplications in this operation to the entire MHA ResBlock can be roughly calculated as follows:
\begin{equation}
\begin{split}
\begin{aligned}
&\frac{s^264^2h}{s^264^2 h+3(64s(d_{model})^2 )h + s(d_{model})^3 + (64s^3)h}\\
&=\frac{s}{s+256h^2+64}.
\end{aligned}
\end{split}
\end{equation}
Since $256h^2$ is no smaller than 16,384 and $s$ is usually no bigger than 128, this ratio should be very small, which illustrates that the management of this single operation will not influence the overall hardware utilization much. If $s$ is smaller than 64, it can be done with the $s\times64$ SA through zero padding to the $K_i$. Otherwise by partitioning the $Q_i$, the $s\times64$ SA can still support this operation with little impact on the utilization of the SA.

\begin{algorithm}
    \SetAlgoLined
    \caption{The Overall Computation Flow}
    \label{alg:System}
    \If{Calculating MHA ResBlock}
    {
        \For{$i=1; i \le h; i++$}
        {
            Temp1=$QW_{Qi}+Bias_{Qi}$\;
            Temp2=$KW_{Ki}+Bias_{Ki}$\;
            Softmax Input=$Temp1\times Temp2^T$\;
            Temp1=Softmax output, Temp2=$VW_{Vi}+Bias_{Vi}$\;
            $P_i$=$Temp1\times Temp2$\;
        }
        \For{$i=1; i \le h; i++$}
        {
            $G_i$=P$\times W_{Gi}+Bias_{Gi}+Q_i$\;
        }
        Output=LayerNorm(G)\;
    }
    \If{Calculating FFN ResBlock}
    {
        \For{$i=1; i \le 4h; i++$}
        {
            $P_i$=ReLU$(XW_{1_i}+b_{1_i})$\;
        }
        \For{$i=1; i \le h; i++$}
        {
            $G_i$=$PW_{2_i}+b_{2_i}+X_i$\;
        }
        Output=LayerNorm(G)\;
    }
    \Return{Output}

\end{algorithm}

\section{Hardware Architecture Design for the Proposed Accelerator}
\begin{figure}[]
\vspace{-0.1cm}
    \centering
    \includegraphics[scale=0.35]{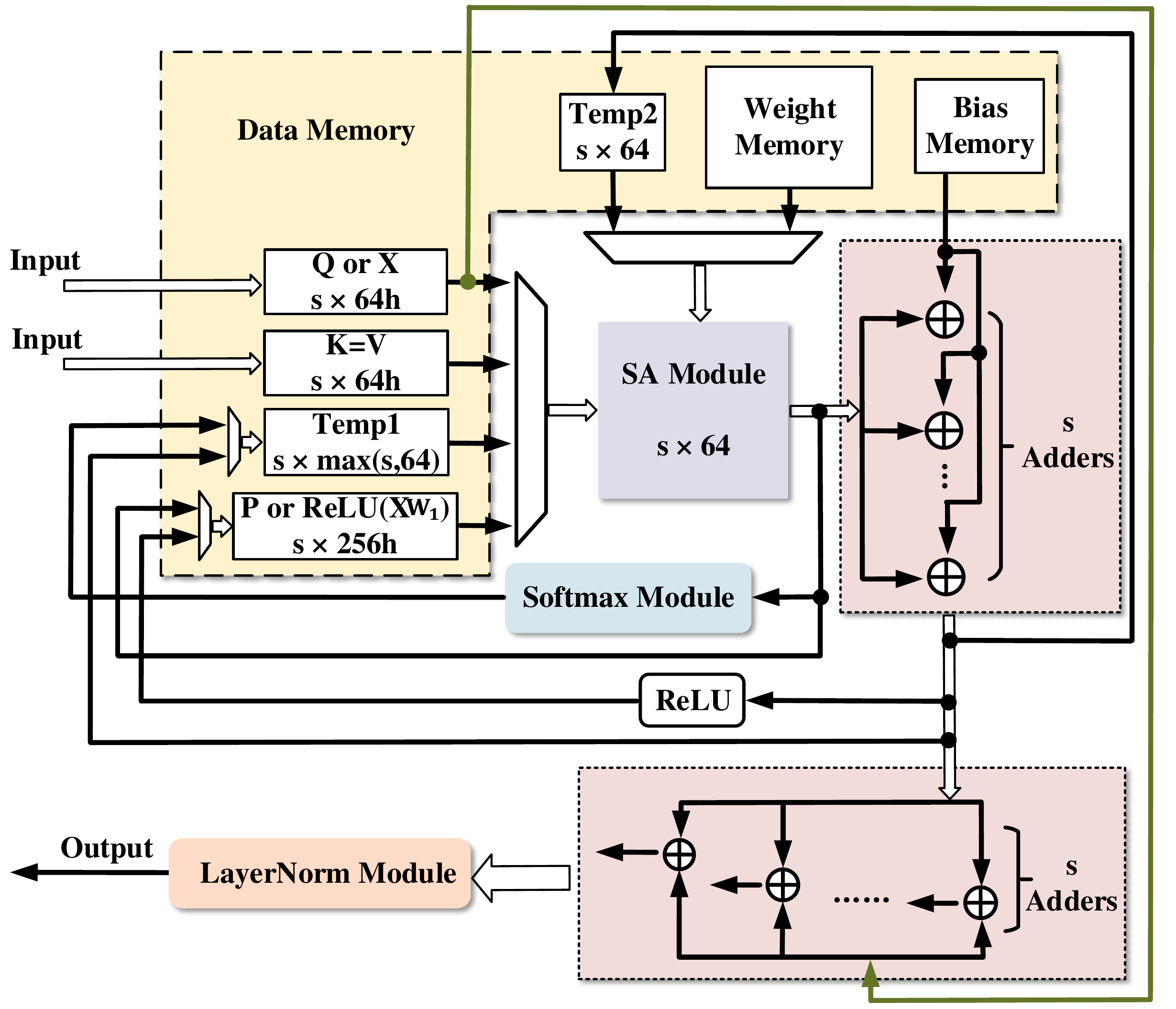}
    \vspace{-0.2cm}
    \caption{The top-level architecture of our design.}
    \label{fig:overview}
\vspace{-0.3cm}
\end{figure}

Using the proposed method of partitioning these weight matrices, the complete hardware accelerator is designed. The top-level architecture is illustrated in Fig.~\ref{fig:overview}.

The $s\times64$ SA is made up of a 2D array of processing elements (PE), with $s$ rows and 64 columns. It is designed to output the product matrix column by column, so each column has $s$ elements. Connected to the SA output, $s$ adders are required to add the bias to the product matrix, and another $s$ adders are required to add the residual before calculating the layer normalization function.
Overall, the SA Module has the highest computational complexity, containing at least $64s$ multipliers and $64s$ adders. To increase the hardware utilization, we make the calculations of the Softmax Module running parallel to $V\times W_{v_i}+Bias_{V1}$ (line 6 in Algorithm \ref{alg:System}). 
Owing to carefully designing the computation flow of the entire system, the SA Module will hardly stop running until the LayerNorm Module starts. As long as the Softmax module can give the output no later than the SA module finishing calculating ``$VW_{Vi}+Bias_{Vi}$'', the latency of the entire system will be determined by the SA module and the LayerNorm module. The architectures of these two nonlinear modules are introduced in detail as follows.

\begin{figure}[]
    \centering
    \includegraphics[scale=0.35]{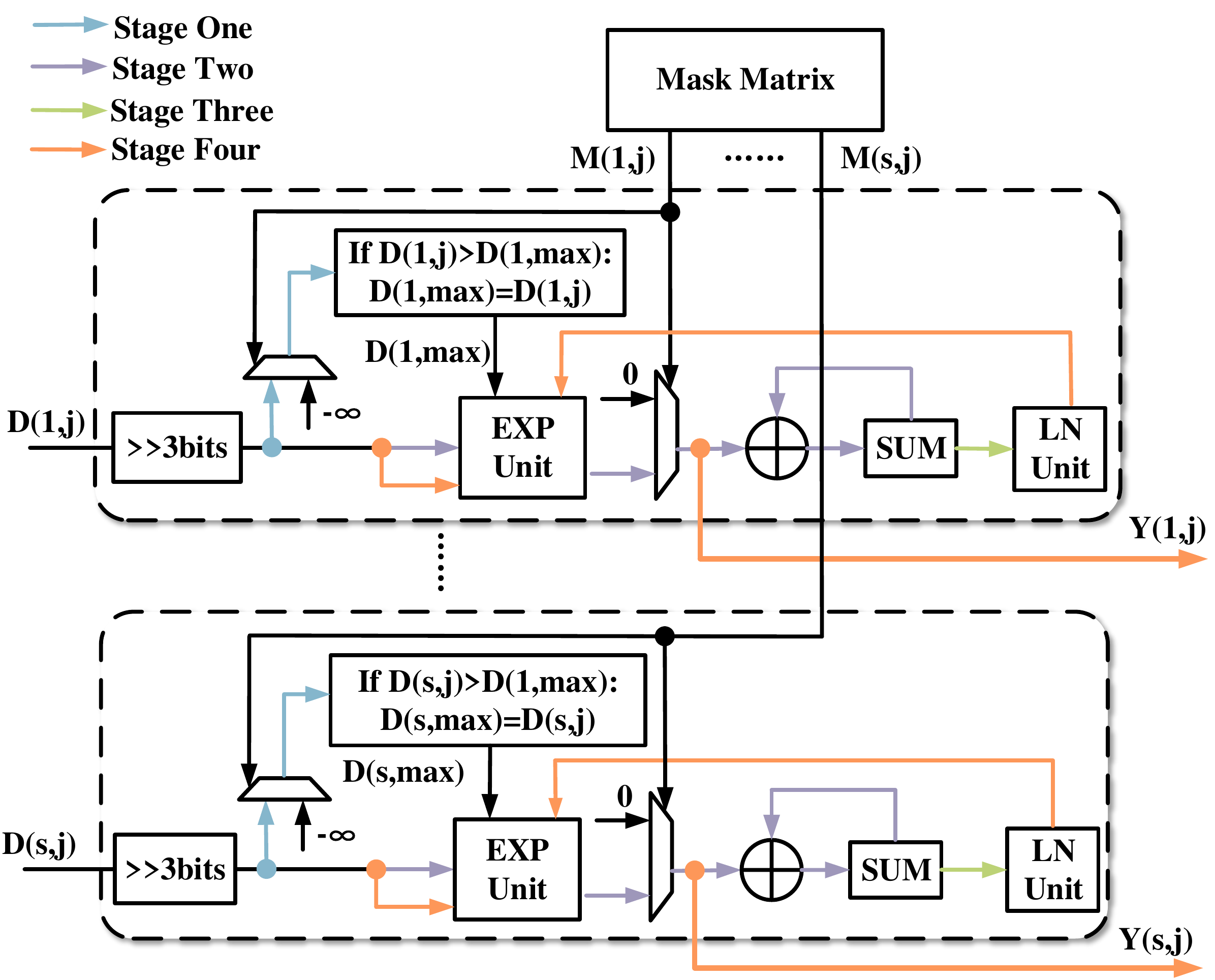}
    \vspace{-0.2cm}
    \caption{The architecture of Softmax module. The ``$>>$'' denotes right shift operation.}
    \label{fig:Softmax}
    \vspace{-0.3cm}
\end{figure}

\subsection{Scaled Masked-Softmax}
The Softmax module in the proposed architecture is used to calculate the scaled masked-softmax function.
For the convenience of discussion, we named
the input matrix $Q_i\times K_i^T$ (refer to line 5 in Algorithm \ref{alg:System}) as $D$, the shape of which is $s\times s$. The output matrix is defined as $Y$, and the mask matrix is defined as $M$. Therefore, the scaled masked-softmax function can be expressed as:
\begin{equation}
\begin{split}
\begin{aligned}
&~~~Y(i,j)=\\
&
\begin{cases}
exp(\frac{D(i,j)}{8})/\sum_{j=1,M(i,j)=0}^s(exp(\frac{D(i,j)}{8}))& M(i,j)=0,\\
0 & M(i,j)=1.
\end{cases}
\end{aligned}
\end{split}
\end{equation}

Although the computational complexity of this Softmax Module is lower than the SA module, the exponentiation and division calculations are still quite expensive.
In \cite{wang2018high}, by making good use of the log sum-exp trick\cite{Yuan2017Efficient} and designing algorithmic reduction strategies for exponential function and logarithmic function, a high-speed and low-complexity hardware architecture for softmax function was proposed. These tricks and strategies are also used in this work to build an efficient architecture for scaled masked-softmax.
The division calculation and numerical underflow can be avoided by using the log-sum-exp trick ($\forall i \in {1,2,...,d_k}, \chi_{max}\geq \chi_i$):

\begin{equation}\label{eq:log-sum-exp}
	\begin{split}
		\begin{aligned}
			Soft&max(\chi_i) = \frac{exp(\chi_i-\chi_{max})}{\sum_{j=1}^{d_k}exp(\chi_j-\chi_{max})} \\
				&= exp(\chi_i-\chi_{max} - ln(\sum_{j=1}^{d_k}exp(\chi_j-\chi_{max}))) \\
		\end{aligned}
	\end{split}
\end{equation}

According to Equation (5), the computation of this module can be broken into four different phases, which is described in Fig.~\ref{fig:Softmax}.
The transformations of exponential function and logarithmic function allow us to build the Softmax module without using any regular multipliers and lookup tables.
The detailed architectures of the EXP Unit and the LN Unit are the same as \cite{wang2018high}.

\subsection{Layer Normalization}
\begin{figure}[htbp]
\vspace{-0.3cm}
\centerline{\includegraphics[scale=0.64]{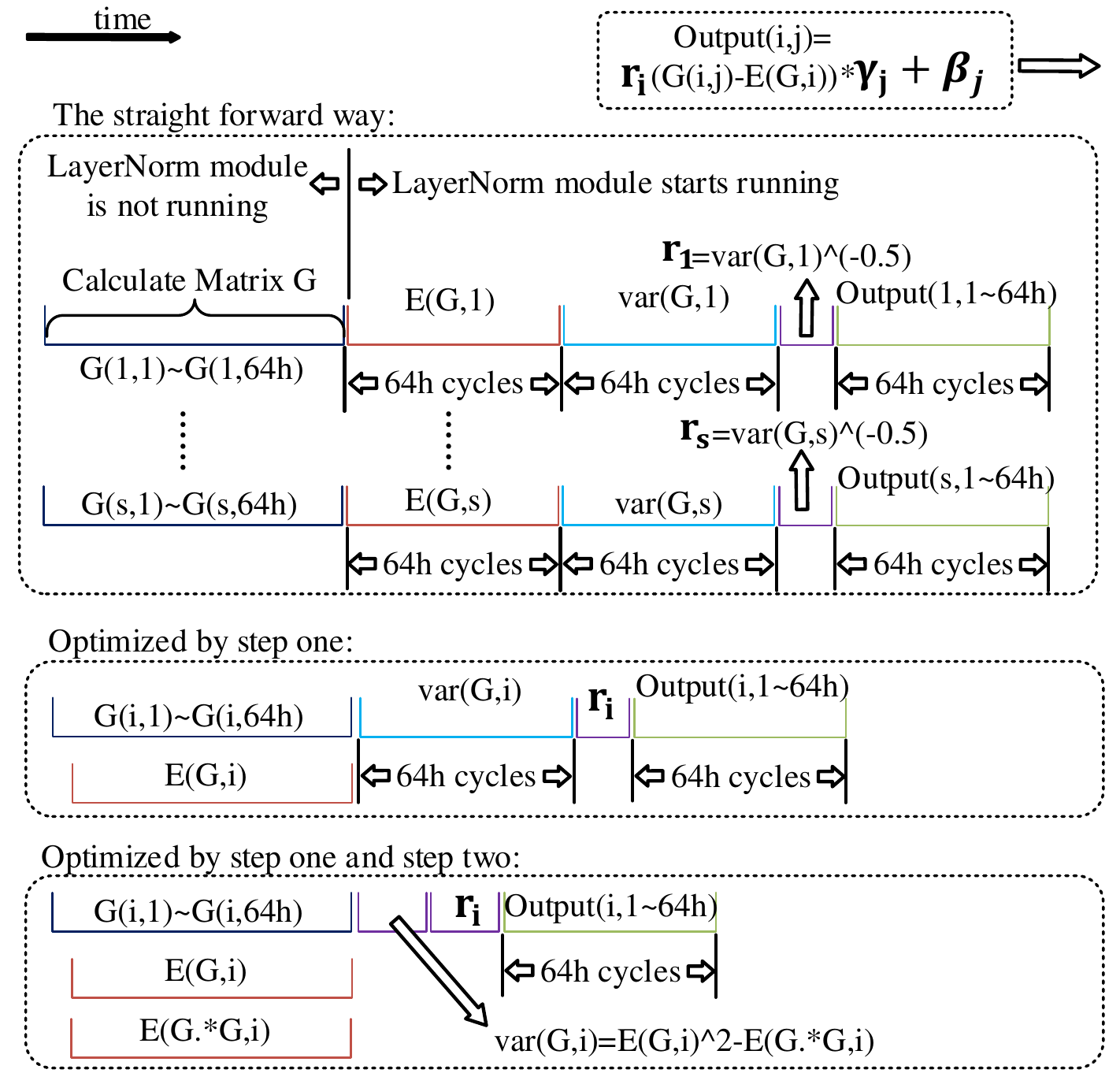}}
\vspace{-0.1cm}
\caption{The method to minimize the latency of the LayerNorm module.}
\label{fig:layernormtiming}
\vspace{-0.2cm}
\end{figure}
As discussed in Section \uppercase\expandafter{\romannumeral2}, both of these two ResBlock have to calculate the layer normalization function before the output starts. This means that the LayerNorm module is always on the critical path of the system latency. In this subsection, we propose a method to minimize its latency.

Unlike the batch normalization, the layer normalization does not impose any restriction on the size of a mini-batch. So it is able to be used in the pure online regime with the batch size equal to 1.\cite{ba2016Layer}. The layer normalization function used in these two ResBlocks is:
 \begin{equation}
Output(i,j)=\frac{G(i,j)-E(G,i)}{\sqrt{var(G,i)+\varepsilon}}\gamma_j+\beta_j,
\end{equation}
where the constant $\varepsilon$ is equal to $10^{-8}$, which is used to avoid the denominator from being zero. The variable $E(G,i)$ is the mean value of all the elements in the $i$-th row of matrix G ($s\times d_{model}$):
\begin{equation}
E(G,i)=\frac{1}{d_{model}}\sum_{k=1}^{d_{model}}G(i,k).
\end{equation}
The variance of these elements is defined as:
\begin{equation}
var(G,i)=\frac{1}{d_{model}}\sum_{k=1}^{d_{model}}[(G(i,k)-E(G,i))^2].
\end{equation}
According to these above equations, the straightforward way to calculate the layer normalization is described in Fig.~\ref{fig:layernormtiming}. To calculate $E(G)$ and $var(G)$, at least $128h$ cycles are added to the whole system latency.

\begin{figure}[]
\vspace{-0.1cm}
    \centering
    \includegraphics[scale=0.35]{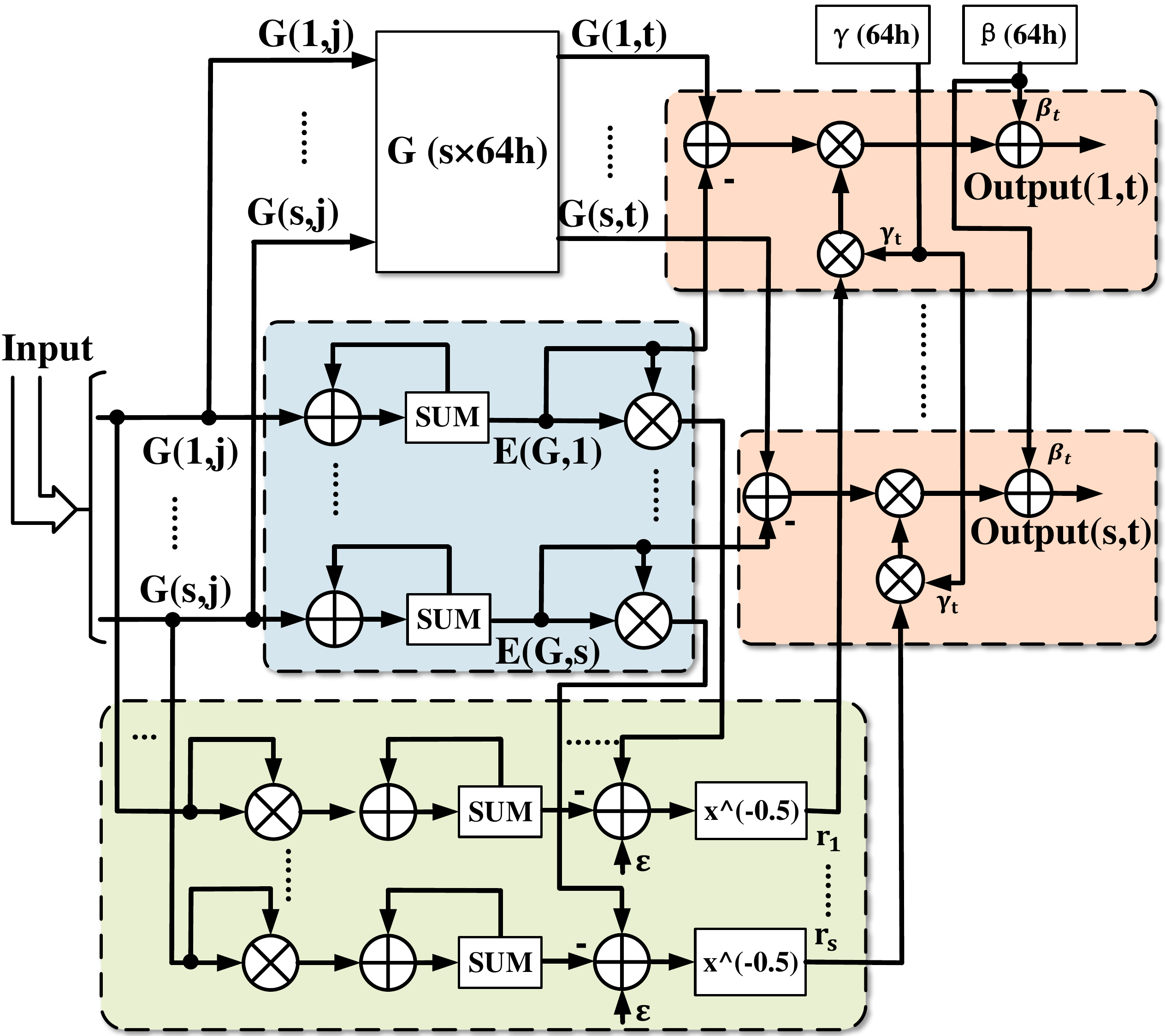}
     \vspace{-0.4cm}
    \caption{The architecture of LayerNorm module.}
    \label{fig:LayerNorm}
    \vspace{-0.3cm}
\end{figure}

As is shown in Fig.~\ref{fig:layernormtiming}, there are two steps in our method of minimizing the delay of this module, and the key is to make the LayerNorm module start running in advance. The first step is using $s$ accumulators to calculate $\sum_{k=1}^{d_{model}}G(i,k)$, and keeping them connected directly to the input of this module.
The second step is choosing another way to calculate the variance:
\begin{equation}
var(G,i)=E(G,i)^2-\frac{1}{d_{model}}\sum_{k=1}^{d_{model}}G(i,k)^2.
\end{equation}
At last, very few cycles are required between the system finishing calculating all the elements of matrix G and starting the output, which also means the latency of the entire system is further reduced.
The architecture of the LayerNorm module is described in Fig.~\ref{fig:LayerNorm}. The ``x$\hat{~}$(-0.5)'' unit is implemented with a lookup table in our experiment.

\section{Experimental Results}
\subsection{Quantization of Transformer Base Model}
Before evaluating our complete design with FGPA, we quantize a Transformer base model for a machine translation task $\footnote{https://github.com/Kyubyong/transformer}$.
This model has been trained and tested with IWSLT 2016 German-English parallel corpus, and the test BLEU score is 23.88 on ``tst2014''. Learning from \cite{bhandare2019efficient}, replacing FP32 with INT8 in the Transformer can greatly reduce the computational complexity with limited accuracy loss.

Since linear approximation is used in the  exponential function and the logarithmic function of the Softmax module, the process of the quantization is divided into two steps. First, all the trainable variable matrices and activation matrices in Fig.~\ref{fig:Matrix} are all quantized with INT8, but the internal calculations in the Scaled Masked-Softmax operation are still implemented with FP32. After that the BLEU score drops to 23.48, proving that quantizing with INT8 in this network is acceptable. Second, the Softmax module is quantized based on the fixed-point model built in the first step. The previously mentioned log-sum-exp trick and the transformations of exponential function and logarithmic function are used. The final BLEU score of the quantized Transformer base model is 23.57, which is even a little higher than 23.48. These results also show that using the simplified architecture for softmax designed in \cite{wang2018high} has little impact on the accuracy of this translation task.

\subsection{Hardware Implementation Results}
By setting the batch size to 1 and the max sequence length to 64, the proposed architecture is evaluated on Xilinx xcvu13p-fhga2104-3-e FPGA by using the Vivado 2018.2. The simulation results show that it takes 21,344 cycles and 42,099 cycles to finish the calculation of MHA ResBlock and FFN ResBlock, respectively. The Vivado implementation results show that our design can run up to 200MHz, and the total on-chip power is 16.7W (13.3W dynamic power and 3.4W device static power). The utilization report is presented in TABLE \ref{Utilization}.

\begin{table}[htbp]
\caption{Utilization Report for the Proposed Hardware Accelerator and its Primary Modules}
\centering
\begin{threeparttable}
\begin{tabular}{c|c|c|c|c}
\hline
\hline
 &\textbf{LUT} & \textbf{CLB Registers}& \textbf{BRAM}& \textbf{DSP} \\
\hline
\textbf{Available}& 1728000& 3456000&  2688& 12288 \\
\hline
\hline
\textbf{Top}& 471563& 217859&  498& 129 \\
\hline
\textbf{64$\times$64 SA}& 420867& 173110&  0& 0 \\
\hline
\textbf{Softmax}& 21190& 32623& 0& 0 \\
\hline
\textbf{LayerNorm}& 10551& 5325& 27.5& 129 \\
\hline
\textbf{Weight Memory}& 3379& 80& 456& 0\\
\hline
\hline
\end{tabular}
\label{Utilization}
\end{threeparttable}
\end{table}

Using the same hyper parameters (batch size equal to 1 and max sequence length equal to 64), we also measure the latency of these two layers in a GPU implementation of the Transformer base model $\footnote{https://github.com/jadore801120/attention-is-all-you-need-pytorch}$ on an NVIDIA V100. The comparison results are shown in TABLE \ref{Comparisons}, proving that our design is able to accelerate the inference for the Transformer on FPGA platform.
\begin{table}[htbp]
\vspace{-0.2cm}
    \caption{Comparisons between FPGA and GPU Latency Results}
    \centering
    \begin{threeparttable}
    \begin{tabular}{c|c|c|c}
    \hline
    \hline
    & \textbf{FPGA Latency} &\textbf{GPU Latency} &\textbf{Speed-Up}\\
    \hline
    \hline
    MHA ResBlock & 106.7us & 1557.8us &14.6$\times$\\
    \hline
    FFN ResBlock & 210.5us & 713.4us  &3.4$\times$\\
    \hline
    \hline
    \end{tabular}
    \label{Comparisons}
    \end{threeparttable}
    \vspace{-0.4cm}
\end{table}

\section{Conclusion and Future Work}
In this work, we present the first hardware accelerator for the MHA ResBlock and the FFN ResBlock in the Transformer. The FPGA implementation shows promising results in terms of both speed and power, which demonstrates that this design can contribute to operating the Transformer network in mobile device or embedded systems. In the future, we will build a FPGA or ASIC accelerator for the complete Transformer inference.


\bibliographystyle{plain}
\bibliography{bibitem}

\end{document}